# Integrated lithium niobate microwave photonics: Driving next-generation wireless technologies

Hanke Feng[1*], Yuansong Zeng[1], Kaixuan Ye[2], Yuansheng Tao[1], Zihan Tao[3], Tong Ge[1,4], Haowen Shu[3,5], Xingjun Wang[3,5,6], David Marpaung[2], Cheng Wang[1*]

[1]*Department of Electrical Engineering & State Key Laboratory of Terahertz and Millimeter Waves, City University of Hong Kong, Kowloon, Hong Kong, China*

[2]*MESA+ Institute of Nanotechnology, University of Twente, Enschede, The Netherlands*

[3]*State Key Laboratory of Photonics and Communications, School of Electronics, Peking University, Beijing, China*

[4]*John A. Paulson School of Engineering and Applied Sciences, Harvard University, Cambridge, MA 02138, USA*

[5]*Frontiers Science Center for Nano-optoelectronics, Peking University, Beijing, China*

[6]*Peking University Yangtze Delta Institute of Optoelectronics, Nantong, China*

[*]*Corresponding authors: hankefeng2@cityu.edu.hk, cwang257@cityu.edu.hk*

**Abstract:** Integrated microwave photonics (MWP) offers a powerful paradigm for handling high-speed microwave signals within chip-scale optical systems. It provides a cost-effective solution to address bandwidth, tunability, and loss bottlenecks of electronics-based radio frequency (RF) systems. The recently emerged thin-film lithium niobate (TFLN) photonic platform, with its exceptional electro-optic (EO) properties, low loss, and scalability, has shown promise to reshape the MWP landscape. Here, we discuss the performance implications of state-of-the-art TFLN photonic devices for MWP applications and offer insights into the emerging trends for next-generation wireless networks. In particular, the unparalleled EO bandwidth enables direct optical generation, processing, and reception of millimeter-wave or even terahertz (THz) signals, significantly expanding the operation frequency range of MWP systems. The low drive voltages and linearity of TFLN modulators lead to an unprecedented operation regime of radio-over-fiber (RoF) systems, featuring net gain, low noise figure and large dynamic range, simultaneously. The availability of a versatile device toolkit, combined with low optical loss and scalability, further supports the transition from traditional tabletop MWP systems to chip-scale solutions, with advanced functionalities, compact footprint, and enhanced system robustness. As the TFLN industrial ecosystem rapidly matures, TFLN-based MWP technology has the potential to deliver transformative solutions to future 6G integrated sensing and communication networks.

## Main

Wireless communication has profoundly transformed modern society over the past four decades. From Bell Labs' initial development of the Advanced Mobile Phone System (AMPS) to the global rollout of the fifth generation (5G) cellular networks, each generation has been driven by ever-increasing demand for higher data capacity, lower latency, and denser connectivity. The forthcoming 6G networks are expected to deliver not only increased data rates, but precise environmental sensing capability, enabling Integrated Sensing and Communications (ISAC) [1-3]. The goal for terabit-per-second connectivity and millimeter-level sensing resolution is pushing carrier frequencies in 6G and beyond networks toward millimeter-wave or even terahertz (THz) regimes [4].

The drive to operate at these ultrahigh frequencies has pushed the limits of traditional electronics and prompted remarkable strides in advanced silicon-based CMOS technologies [5-7]. Yet fundamental challenges remain: key components and interconnects become substantially more lossy and inefficient in upper frequency bands, presenting a bottleneck that is difficult to overcome with electronics alone. As a result, modern radio access networks increasingly employ optical fibers instead of copper cables to distribute signals between centralized or distributed units (CU/DU) and remote radio units (RRU), broadly known as radio over fiber (RoF) [8]. This represents a canonical application of microwave photonics (MWP) technology, which circumvents the inherent bandwidth bottlenecks of electronics by transmitting high-speed microwave signals in the optical domain. Over the past two decades, the application realm of this MWP concept has been significantly expanded, covering more advanced tasks such as microwave generation, processing, and sensing [9,10]. Recent advances in photonic integration technologies have further miniaturized MWP systems to chip scales, termed integrated MWP [11]. To date, most integrated MWP circuits have been implemented on silicon (Si) [12,13], silicon nitride (SiN) [14,15], and indium phosphide (InP) [16,17] platforms, due to the availability of mature and scalable manufacturing ecosystems. Despite significant milestones, these platforms see increasing challenges in meeting the stringent bandwidth, loss and scalability requirements of future 6G systems [18]. Specifically, they necessitate broadband and linear electro-optic modulators capable of efficient signal conversion across the entire wireless spectrum (from sub-6 GHz to THz band), a challenge for Si-based platform due to its inherently limited free-carrier-based modulation mechanism [19]. The vision of adaptive 6G networks demands low-loss, large-scale photonic circuits for programmable signal processing, where InP platform faces critical bottlenecks [11]. Furthermore, the need for minimized system size and power consumption requires seamless co-integration of multiple photonic functionalities on a single chip.

The recently emerged thin-film lithium niobate (TFLN) platform provides a promising solution to these demands [20,21]. Compared to legendary bulk LN devices, the TFLN platform retains nearly all the advantageous material properties, including excellent EO proprieties, power handling ability and broad transparency window, while strongly confining light in a small mode volume with enhanced light-matter interaction. Figure 1a summarizes the performance metrics critical to MWP applications, including modulation efficiency (voltage-length product), bandwidth, propagation loss, power handling, and scalability, for popular photonic platforms. TFLN clearly stands out as an all-round choice with favorable properties that are not simultaneously available on other photonic platforms. This synergy enables record-breaking device-level building blocks for integrated MWP, including EO modulators [22-34] with unprecedented bandwidths, CMOS-compatible driving voltages, and ultrahigh linearity; versatile signal processing toolkits with low loss and reconfigurability [35-38]; as well as highly flexible and broadband optical frequency combs [39-45]. Moreover, the rapid maturation of the TFLN industrial ecosystem, with 8/12-inch wafer readiness [46], is accelerating the integration of these high-performance building blocks into large-scale manufacturable photonic integrated circuits with enhanced cost-effectiveness and commercial viability. With this foundation of performance and scalability in place, TFLN photonics is poised to push the performance boundaries of MWP systems toward previously inaccessible regimes, unlocking emerging application scenarios in wireless communications and sensing.

In this Perspective, we provide insights and analysis on the exciting new regimes that TFLN photonics are driving MWP systems into, including (i) RoF link with net gain, (ii) ultra-broadband signal transmitter covering mmWave and THz bands, (iii) wireless receiver with on-chip antenna integration and low noise figures, and (iv) ultra-fast programmable signal processing (Fig. 1a). The broadband, multi-functional and scalable TFLN MWP infrastructure could ultimately provide unified, compact, and cost-effective solutions for emerging 6G ISAC applications, including Vehicle-to-Everything, THz communication/radar, and Immersed Reality (Fig. 1b).

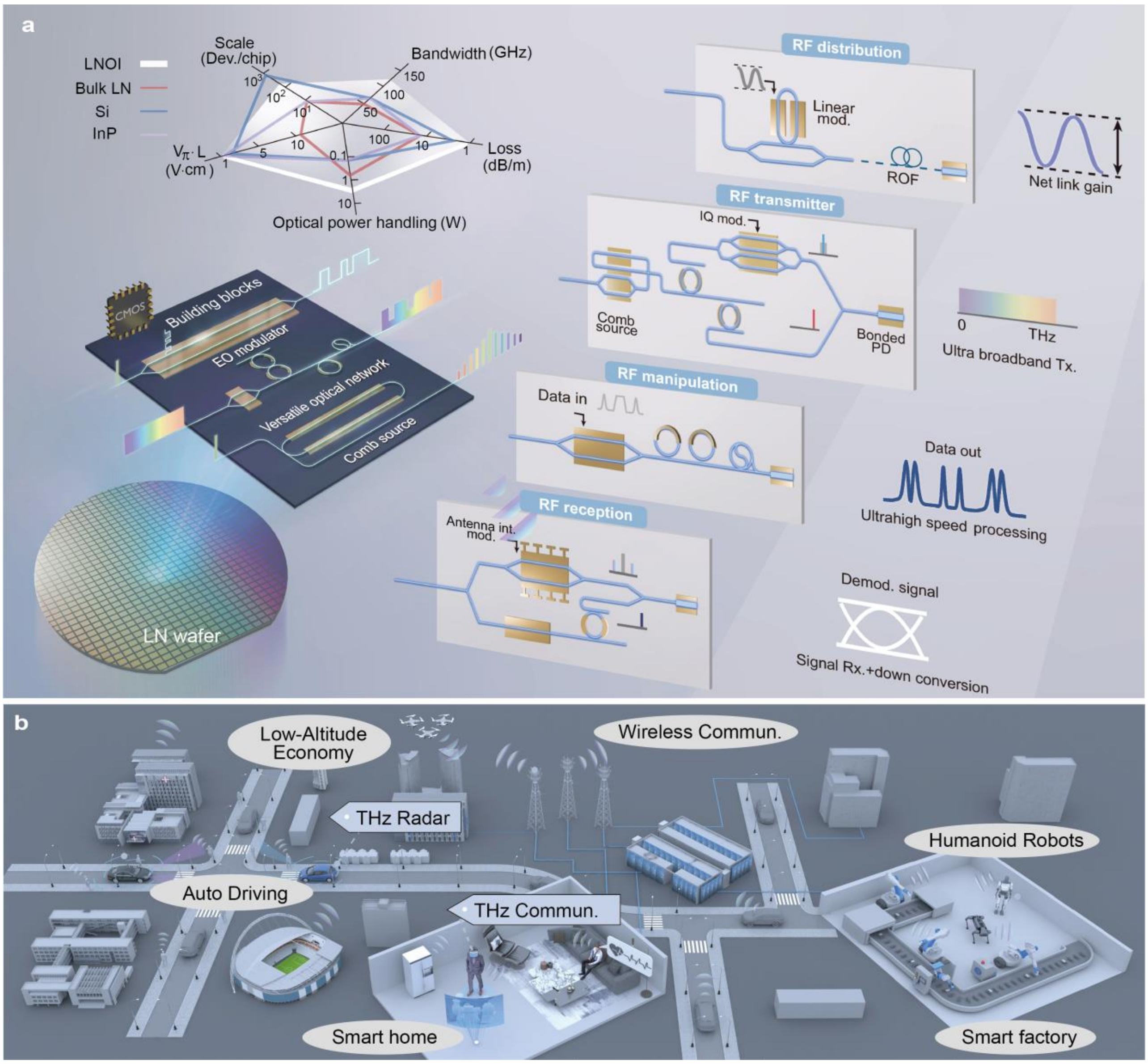

**Fig. 1. TFLN MWP systems and envisioned application scenarios. a.** Performance pentagon of different MWP platforms (Top left). Versatile TFLN MWP building blocks, including high performance electro-optic modulators, low-loss optical networks, and broadband comb sources are manufacturable on wafer scales, enabling on-chip MWP systems with unprecedented performances, including RF distribution with net link gain, RF transmitter covering THz band, RF manipulation with ultrafast speed, and RF reception with antenna integration. **b.** Envisioned next-generation application

scenarios empowered by TFLN MWP technologies. Mod., modulators; IQ, in-phase and quadrature; Tx., transmitter; Rx., receiver; RoF, radio over fiber; PD, photodetector.

**RF transmission: Linear RoF links with net gain and low noise figure.**

Despite being a key enabling technology for modern radio access networks, current RoF links suffer from limitations such as severe link losses typically > 20 dB (the power ratio between output and input RF signals), on top of nonlinearity and noise issues [10]. The root cause lies in the limited conversion efficiencies of the electro-optic (E/O) and opto-electronic (O/E) conversion processes. These bottlenecks fundamentally constrain the performance of RoF links, since link losses are accompanied by elevated noise figures (loss of signal-to-noise ratios during this process) that cannot be recovered by subsequent amplifiers.

The advances of TFLN technology offer unique opportunities to drive RoF links beyond their traditional performance boundaries, with net link gain and low noise figure simultaneously (Fig. 2a). On the EO conversion front, TFLN modulators provide significantly lower modulation half-wave voltages ($V_\pi$), especially at high frequencies, than their bulk counterparts, leading to increased EO conversion efficiency. Figure 2b summarizes the $V_\pi$ values of state-of-the-art TFLN modulators based on Si (green) and quartz (purple) substrates, in comparison with those of bulk LN modulators (red). The reported values are closely bounded by our estimated performance envelope (blue dashed curve), which is calculated based on the typical conductor loss of 0.2 dB $cm^{-1}GHz^{-1/2}$ and dielectric loss of 0.02 dB $cm^{-1}GHz^{-1}$ (for frequencies > 100 GHz) of the microwave transmission lines [22]. On the other hand, the O/E conversion efficiency is mainly determined by the optical power received at the photodetector (PD). The high power-handling capability (> 30 dBm) and low insertion loss (< 3 dB) of TFLN modulators [22] open up the possibility of saturating high-power PD without post-modulation amplifiers that often lead to deteriorated noise figures.

To validate this prospect, our preliminary link measurements, obtained using a TFLN modulator with an ultralow DC $V_\pi$ of 0.64 V and a high-power handling PD (27 dBm), demonstrate net link gain of +32 dB and low noise figure of 14 dB at 1 GHz (Fig. 2c). The link gain and noise performances degrade at increased frequencies due to efficiency roll-off of both the modulator and the PD, nevertheless net link gain is achieved at frequencies up to 20 GHz. Based on the state-of-the-art modulator performance envelope in Fig. 2b, we predict that even higher gain could be achieved (blue dashed line in Fig. 2c), including > 25 dB gain at 20 GHz and net gain up to 40 GHz. Further incorporating low-loss fiber-chip couplers, which are now commonly employed in the TFLN platform [47,48], could eliminate the need for EDFA to saturate the PD, leading to further decreased system noise figure (red dashed line in Fig. 2c) by mitigating amplified spontaneous emission (ASE) noise.

Besides gain and noise performances, the high power-handling capability of TFLN modulators is also crucial to achieving RoF links with high linearity, an important requirement for analog signal transmission. By increasing the optical power received at the PD from 10 dBm to 23 dBm, our measured spurious-free dynamic range (SFDR) [10] of a simple TFLN Mach-Zehnder interferometer (MZI) modulator significantly increases from 100 $dB \cdot Hz^{2/3}$, a typical value consistent with prior reports limited by the inherent sinusoidal transfer function of MZI [49,50], to 117 $dB \cdot Hz^{2/3}$ (purple triangles in Fig. 2d). This is because an increased optical power concurrently elevates the fundamental and third-order intermodulation (IMD3) components under a fixed noise floor, leading to a higher third-order intercept point and thus a better SFDR performance. Moving forward, the link linearity can be further improved by adopting structural linearization strategies that have proven effective on the TFLN platform [51-55]. For example, combining the high-power PD with a ring-assisted MZI modulator [51], we observe readily increased SFDR to 134 dB·Hz$^{4/5}$ at 1 GHz (green dot in Fig. 2d). A major drawback of such resonator-assisted linearization schemes is the inherent trade-off between linearity and bandwidth (green dashed and solid lines). To achieve broadband linearization, post-modulation processing schemes, e.g. by tailoring the amplitude and phase of the optical carrier using two microring resonators (Fig. 2a), can be employed to achieve frequency-independent suppression of the IMD3 components. Based on previously reported results (blue squares), we expect the SFDR to approach 130 dB across at least 20 GHz bandwidth by utilizing the high-power PD (blue dashed line). Overall, the synergy between low-voltage TFLN modulators and high-power PDs could fuel a new generation of energy-efficient and low-noise RoF infrastructure for 6G and beyond.

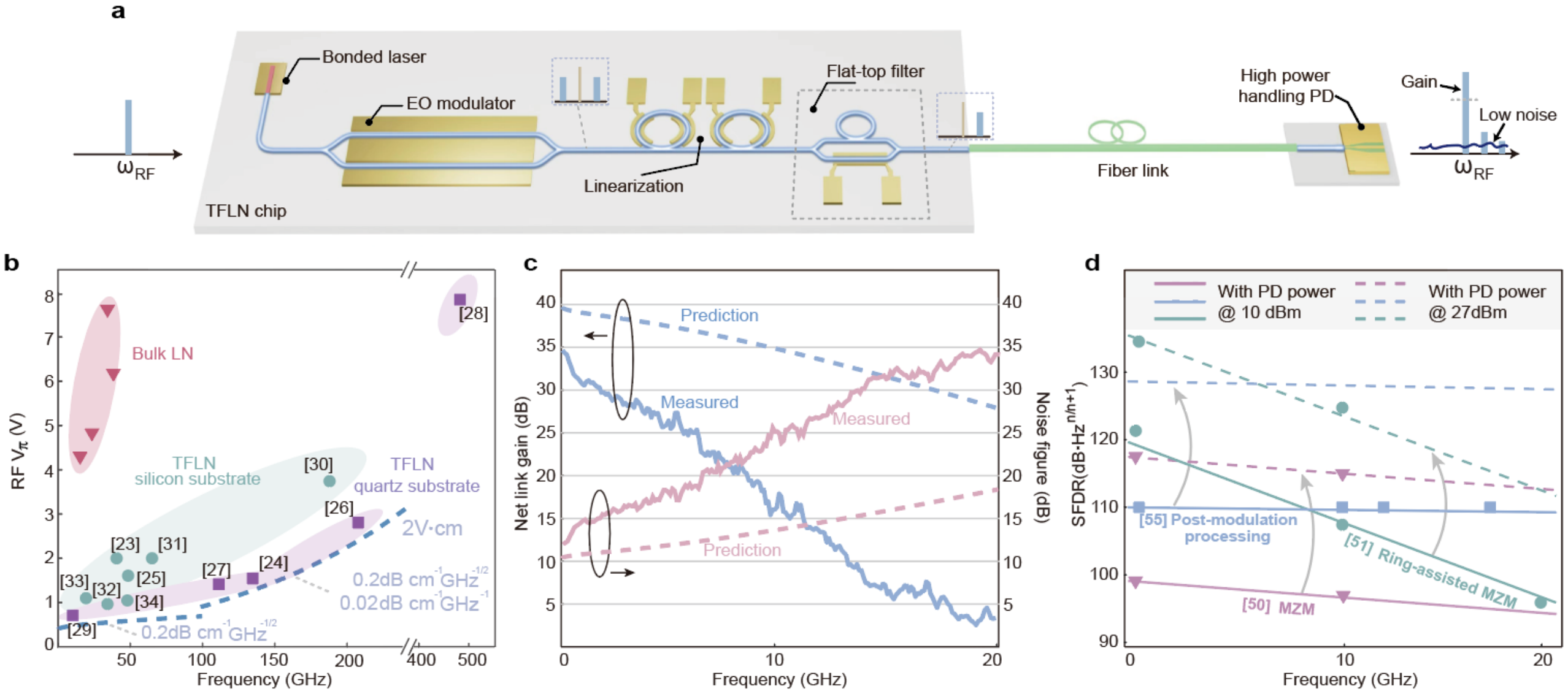

**Fig. 2. Radio-over-fiber links with net gain and low noise figure. a.** Schematic of the net-gain and low-noise MWP link based on highly efficient E/O/E interfaces. **b.** RF half-wave voltage of state-of-the-art TFLN modulators versus frequency, including those based on silicon (green) and quartz (purple) substrates, in comparison with bulk LN modulators (red), with data from [23-33]. Dashed lines represent an estimated performance envelope based on the conductor and dielectric losses of typical transmission line. **c.** Measured and predicted net link gain and noise figures as functions of frequency using a low-voltage TFLN modulator and a high power-handling PD. **d**. Measured SFDR performance versus frequency for various device and circuit designs, including simple MZI modulator (purple triangles)[50], ring-assisted MZI modulator (green dots)[51], and post-modulation processing scheme (blue squares)[55]. Solid and dashed lines correspond to fitted or predicted SFDR values based on normal- and high-power PDs, respectively.

## RF generation: Ultra-broadband low-noise millimeter-wave/terahertz sources

Unlocking the full potential of future mmWave/THz wireless networks requires high-quality RF sources for carrier and local oscillator (LO). Photonic generation techniques provide a compelling alternative to electronic schemes by offering inherent ultra-wide bandwidth, frequency tunability, and low phase noise. The most straightforward photonic approach employs heterodyne beating of two optical tones on a high-speed PD (Fig. 3a, i). In this case, the spectral purity of the generated signal is critically dependent on the phase coherence between the two optical tones. An elegant tool for this purpose is the optical frequency comb, which intrinsically provides a broad range of coherent tones for high-quality signal generation. Following the recent proliferation of Kerr/EO combs on TFLN [39-45], flexible selection of arbitrary comb lines can be straightforwardly achieved using on-chip filters. For example, using a tunable microring filter, a single target comb line can be picked with extinction ratios to neighboring lines exceeding 30 dB (Fig. 3b) [56].

An alternative route for generating broadband and low phase noise signal is through optoelectronic oscillation (OEO) [57], a self-sustained oscillation loop by filtering and amplifying spontaneous noise (Fig. 3a, ii). A distinct feature of OEO is that its oscillation frequency is determined by the filters within the loop, thus can be widely tuned without reliance on external RF sources. Unlike traditional multiplier-based RF generation schemes with cascaded noises at higher frequencies, the phase noises of OEO are generally frequency invariant, and can be significantly suppressed by increasing the optical fiber length for longer cavity lifetime. While numerous OEO schemes have been reported based on discrete components and integrated platforms, the achievable frequency range is usually limited by the modulation bandwidth at the EO interface, typically below 40 GHz [58]. The ultrabroad bandwidth offered by TFLN platform has unleashed the full potential of OEOs, recently pushing the oscillation frequency to 115 GHz while maintaining consistent phase noise around –110 dBc/Hz (with 2km fiber delay) at a 10 kHz offset across the entire band (Fig. 3c) [59].

Notably, both above RF generation schemes naturally allow encoding high-speed data signals on the generated RF signal by electro-optic modulation on the optical carrier. Leveraging the excellent modulation capability and scalability of the TFLN platform, these RF generators can be further integrated with in-phase/quadrature (I/Q) modulators and filters for frequency-tunable, low-noise, and high-speed THz transmitters (Fig. 3a). To realize such fully on-chip THz transmitter architectures, another critical building block is broadband PD that can ultimately convert the generated and processed optical signals back to the RF domain. In this regard, heterogeneous integration of uni-traveling-carrier PD (UTC-PD) provides a compelling solution [60-63], with recent demonstration on the TFLN platform reaching 3-dB bandwidths exceeding 230 GHz [62] (Fig. 3d).

In addition to direct data modulation for communication purposes, other OEO modalities, such as time- or Fourier-domain mode lock schemes [64-67], could also be exploited to generate sophisticated pulse/linear frequency-modulated signals, which are significant for radar sensing applications. The versatility for sensing and communication, unified on the TFLN platform, paves the way for ISAC applications in 6G networks.

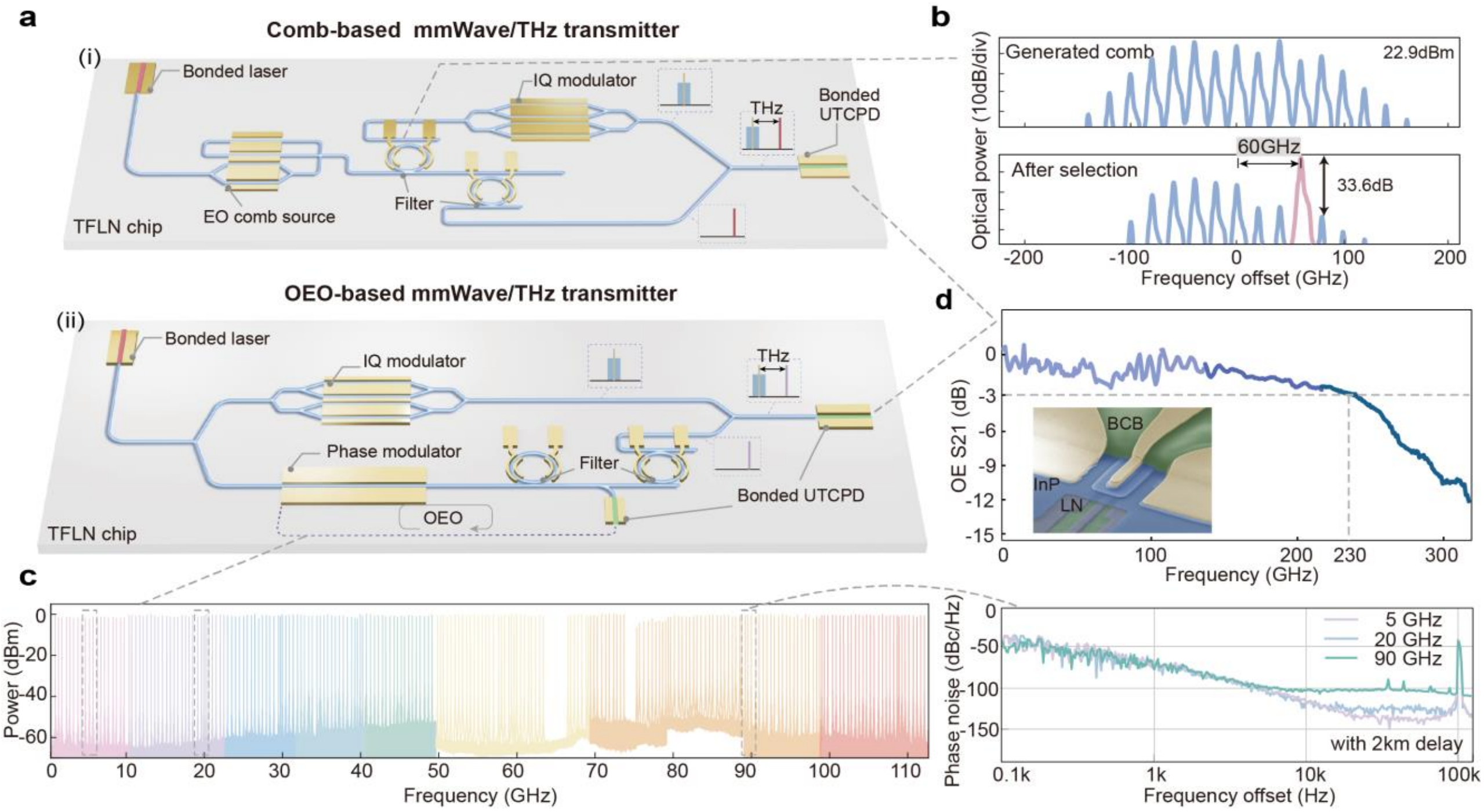

**Fig. 3. Ultra-broadband and low-noise mmWave/THz transmitter. a.** Schematics of integrated mmWave/THz transmitters based on (i) optical heterodyne beating, and (ii) optoelectronic oscillation (OEO) schemes. **b.** Optical spectra of an as-generated (top) and filtered (bottom) electro-optic comb, showing effective selection of a comb line 60 GHz away from carrier using an add-drop microring resonator, with data from [56]. **c**. Measured OEO spectrum from 0.5 GHz to 115 GHz. Right panel shows the phase noises of OEO signals at various frequencies showing consistent performances across bands, with data from [59]. **d.** Frequency response of a bonded UTC-PD on TFLN platform with 3 dB bandwidth up to 230 GHz, with data from [62]. Inset shows the scanning electron microscope image of the bonded UTC-PD.

## RF reception: Antenna-integrated photonic RF frontend

The reception of high-frequency signals typically requires low-noise amplification and down-conversion to an intermediate frequency. Such schemes, relying on mixers and frequency-multiplication-based local oscillators (LO), are often plagued by reduced gain and increased noise at high frequencies. Integrated TFLN MWP technologies offer an alternative route that could circumvent these high-frequency electronic limitations. As illustrated in Fig. 4a, this architecture directly converts the received RF signals into optical domain via broadband TFLN modulators, and performs the subsequent filtering and processing tasks optically. The processed signals can be naturally down-converted to intermediate frequencies by mixing with an optical LO generated using either a filtered comb or OEO, similar to those described in the previous section. This architecture potentially eliminates the need for costly high-frequency electronic devices in the RF frontend.

Following the trends in THz electronics on antenna integration, the MWP receiver can also be further integrated with radiating elements to mitigate the coupling loss and complexity in discrete systems. This strategy is particularly feasible at THz frequencies, where the antenna dimensions, dictated by the reduced wavelength, are highly compatible with the photonic chip scale. Significant progress has been made toward this vision, with the demonstrations of various antenna integrated TFLN modulators tailored for specific operational bands (Fig. 4b) [68-71]. This portfolio of antenna types, such as dipole antenna arrays for efficiency, bowties and slots for broadband operation, and leaky-wave structures for high directivity, enables the optimization of reception systems across diverse applications. Figure 4c shows our preliminarily measured modulation sidebands of TFLN modulators integrated with distributed dipole antenna arrays designed for various frequency bands, showing effective reception of wireless RF signals near 140 GHz, 270 GHz, and 450 GHz. The performance of these antenna-integrated modulators can be quantitatively evaluated by the carrier-to-sideband conversion ratio (CSR) per unit free-space power density (W/cm$^2$), which reflects the overall RF-to-optical conversion efficiency. The devices achieve consistent figures of merit ~ 10 % per W/cm$^2$ across the three bands [Fig. 4d (i)], showing excellent performance scalability to higher frequencies that are typically not seen in electronic solutions.

The most appealing feature of this antenna-integrated receiver architecture is that it eliminates the need for costly and sometimes inaccessible high-frequency, low-noise amplifiers by moving the amplification stage to intermediate frequencies after O/E conversion, which could be well performed using mature electronic solutions. However, it also means the noise figure of this full E/O/E link is critical, since any loss in signal-to-noise ratio cannot be recovered in later stages. Practically, this architecture only makes sense if the overall link gain (the power ratio between signals at the output intermediate frequencies and received THz frequencies) and noise figure can be comparable with electronic solutions. We evaluate these critical system-level metrics based on the antenna performance achieved in Fig. 4c and the performance envelope of EO modulators in Fig. 2b. Assuming the use of a 2-dB-insertion-loss TFLN chip to avoid EDFA, a high-power balanced PD with a saturation power of 30 dBm, and a mature 41-dB-gain low-noise amplifier with a noise figure of 1.8 dB at intermediate frequency, the predicted link gain is near 40 dB across a broad bandwidth from 100 to 500 GHz, as shown in Fig. 4d (ii). The estimated noise figures [Fig. 4d (iii)] of the full THz photonic receiver are comparable to those of state-of-the-art electronic solutions at below 300 GHz [72-74], and are superior beyond 300 GHz thanks to its excellent frequency scalability, while free from any high-frequency electronic components.

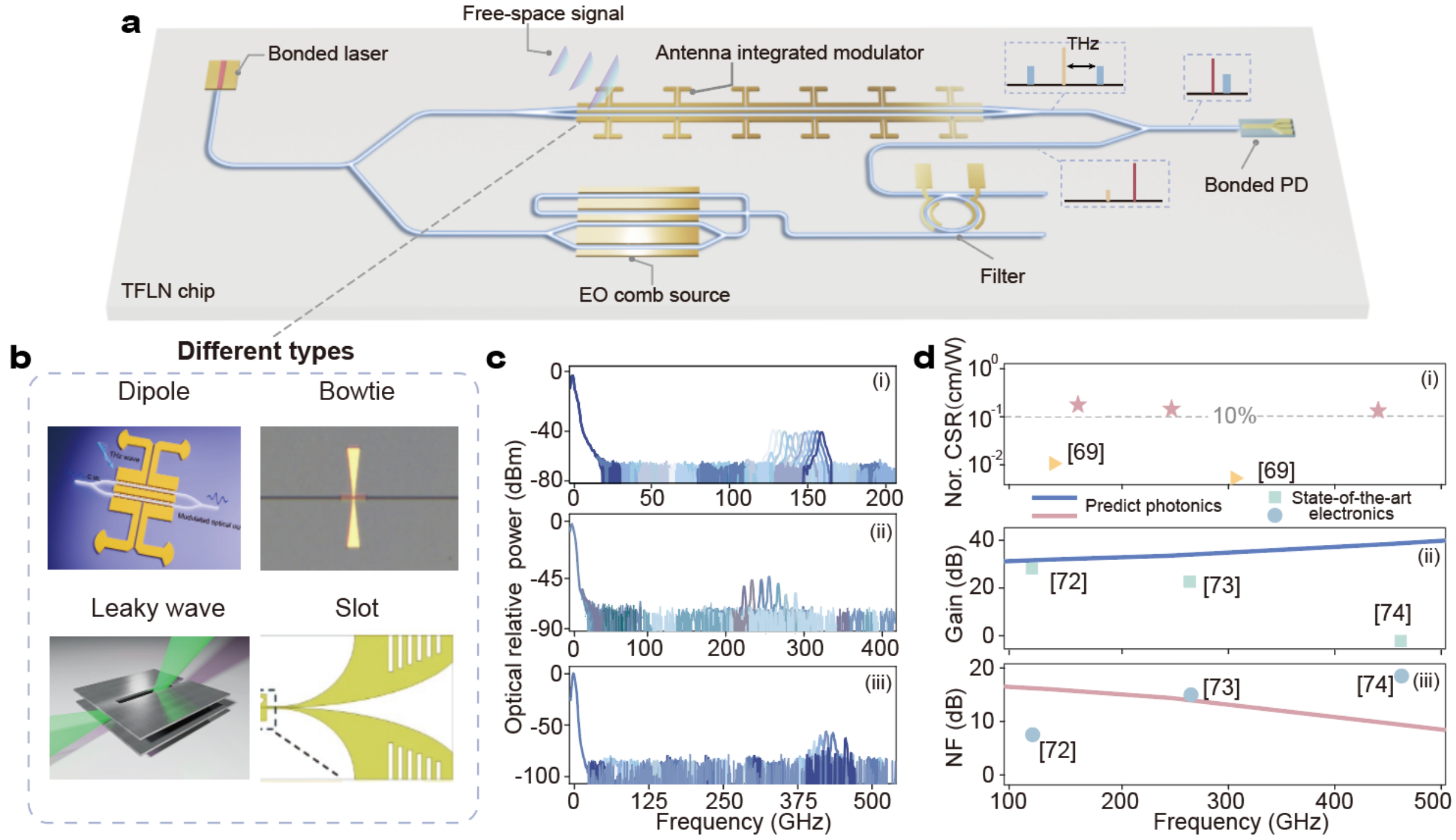

**Fig. 4. Antenna-integrated photonic RF receiver. a.** Schematic of the receiver frontend capable of interception of free-space signals, broadband electro-optic conversion, optical mixing and processing, and finally down-conversion to intermediate frequencies. **b.** Various types of RF antennas integrable with TFLN modulators, including dipole array, bowtie, leaky wave, and slot architectures. **c.** Measured optical spectra of antenna-integrated modulators with input free-space signals near 140 GHz (i), 270 GHz (ii), and 450 GHz (iii), respectively. **d**. Measured carrier-to-sideband ratios (CSR) per unit power density versus frequency (i), estimated link gain (ii) as well as noise figure (iii) of the photonic reception system at various frequencies, in comparison with state-of-the-art electronics[72-74].

## RF processing: Ultra-fast programmable signal processing

Advancing RF signal processing towards on-the-fly operations and programmability is critical for future wireless networks, which however poses significant challenges to electronic circuits due to fundamental bandwidth, latency, and tunability constraints. Migrating such functions to the optical domain, leveraging its nearly infinite bandwidth and frequency reconfigurability, provides a promising solution. To achieve this, an ideal MWP signal processing platform should simultaneously provide high-speed, efficient modulation for EO conversion, low-loss programmable optical processing networks, and large-scale integration capability. TFLN platform fulfills all these requirements (Fig. 5a), as demonstrated in our previous work realizing on-chip EO modulation and processing of analog signals with ultrabroad bandwidths up to 67 GHz [75]. A 9.6-ps Sinc pulse is accurately processed by a temporal differentiator on TFLN chip, showcasing the exceptional processing speed and fidelity (Fig. 5b) [75]. Further incorporating programmable optical processing units, for example cascaded microring filters with tunable coupling strengths and linewidths, have enabled programmable MWP filters [55,76,77] with broad bandwidths up to 60 GHz, high resolutions down to hundreds of MHz, and extinction ratios beyond 80 dB (Fig. 5c) [55]. These versatile and high-performance building blocks make the TFLN platform particularly powerful for demanding real-world applications. A compelling example is photonic radar systems, where TFLN chips have been utilized for the de-chirping of V-band (40–50 GHz) echo signals, enabling precise ranging and velocity measurement with centimeter-level accuracy (Fig. 5d) [78]. Moving forward, the broadband EO modulation unit and scalable signal processing core could be further integrated with EO comb generators, configured either in serial or parallel, for frequency up-/down- conversion, microwave synthesis, and optical sampling, where the comb can serve as multi-wavelength carriers or coherent reference sources (Fig. 5a).

Another important MWP signal-processing mechanism recently proposed and validated on the TFLN platform is stimulated Brillouin scattering (SBS) [79,80]. This breakthrough, enabled by photon-phonon interaction, unlocks ultra-high resolution MWP filters with MHz-level linewidths (Fig. 5e), which are otherwise difficult to achieve using passive optical filters. The bandwidth of the filter can further be tailored by exploiting the anisotropic Brillouin gain spectrum of TFLN, providing unparalleled flexibility [79].

Besides processing and filtering received RF signals, TFLN-based MWP circuit also enables real-time spectrum sensing with unprecedented analysis bandwidths, which is crucial for dynamic spectrum allocation in future ISAC systems. In our recent demonstration based on a frequency-to-time mapping scheme [81-83], real-time RF spectral analysis is achieved at up to 120 GHz with 100 ns latency, enabling on-the-fly spectral reconstruction of broadband unknown signals (Fig. 5f) [81].

The pursuit of adaptive and intelligent ISAC systems, particularly for directional applications in THz communications, demands advanced beamsteering capabilities. To achieve this on-chip functionality, the integration of agile phase shifters [84] or true-time-delay (TTD) [85,86] modules is essential for multi-stream, multi-directional data transmission within a unified TFLN platform. The feasibility of this approach has been validated by TTD-based beamforming, with squint-free beam steering across a wide angular range from -75.42° to 77.58° at 40 GHz (Fig. 5g) [87].

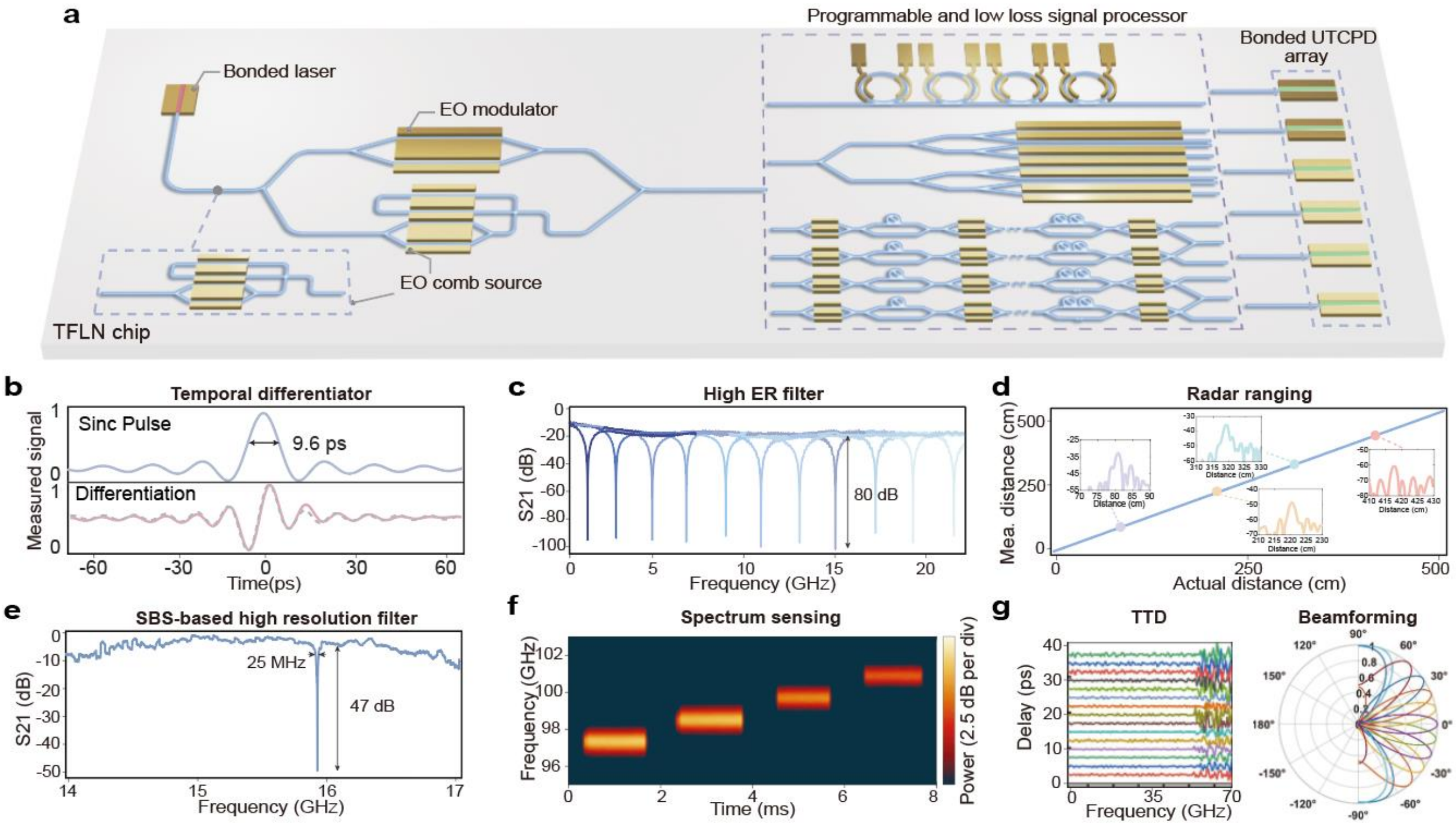

**Fig. 5. Ultra-fast programmable RF signal processing engine. a.** Schematic of versatile RF signal processing on TFLN platform. **b.** Temporal differentiation for an ultrafast Sinc pulse of 9.6 ps, with data from [75]. **c.** Programmable MWP filter with 80 dB extinction ratio, with data from [55] **d**. Measured radar ranging results with distance errors within 0.15 cm, with data from [78]. **e**. SBS-based high-resolution MWP filter with 25 MHz bandwidth, with data from [79]. **f.** Real-time spectrum sensing of a frequency stepping signal around 100 GHz, with data from [81]. **g**. Radiation patterns based on a true-time delay-line with steering angle from −75.42° to 77.58°, with data from [87].

## Outlook: Co-packaged transceivers for 6G ISAC

Recent collective efforts from the TFLN community have set new performance paradigms of MWP systems, giving new life to this established field with transformative possibilities. With bandwidth, tunability, and power efficiency beyond the reach of other integrated photonic platforms, TFLN photonics is on the verge of delivering a variety of MWP functions, ranging from RF generation, transmission, reception, and processing, that can truly outperform electronic solutions in all key performance indicators, as we have discussed and predicted in this perspective.

To fully realize these promises on commercial landscapes, the critical next step is to transition from lab-based proof-of-concept demonstrations based on heroic devices and auxiliary equipment, toward co-packaged full systems with photonic and electronic chips taped out at industrial scales. The rapidly evolving TFLN industrial ecosystem, marked by readiness of 8/12-inch wafers and industry-grade foundries, is emerging as a crucial driver that enables MWP circuits with larger-scale integration, higher yield, and more advanced functionalities. Other necessary functionalities, including lasers, PDs, and peripheral electronics, can be combined with the TFLN core through heterogenous and/or hybrid integration, especially leveraging advanced co-packaged optics (CPO) technologies that have recently achieved commercial maturity. Notably, most architectures envisioned in this perspective only require mature optoelectronic components, e.g. low-speed analog-to-digital converters and digital signal processor, while the more demanding high-frequency operations are addressed by the TFLN core, significantly reducing the cost and complexity in packaging.

Building on these advances, fully co-packaged transceivers with superior performance and versatile functionalities will be ready for deployment in a wide range of 6G ISAC settings (Fig. 6). By selectively integrating key building blocks, including EO comb array, modulator array, beamforming units (phase shifter/TTD modules), bonded UTCPD/laser, optical filter array, as well as antenna-integrated modulator array at transmitter/receiver end, the TFLN photonic core could be flexibly designed to perform high-data-rate wireless transmission and high-resolution sensing across multiple targets with low-latency and software-defined programmability. Such a tightly integrated ISAC engine also offers the capabilities of dynamic spectrum allocation, agile beam steering, and robust interference mitigation under the complex electromagnetic environments anticipated in 6G networks. Altogether, TFLN MWP stands as a foundational hardware engine for next-generation wireless technology, delivering the performance, intelligence, and scalability required for the 6G era and beyond.

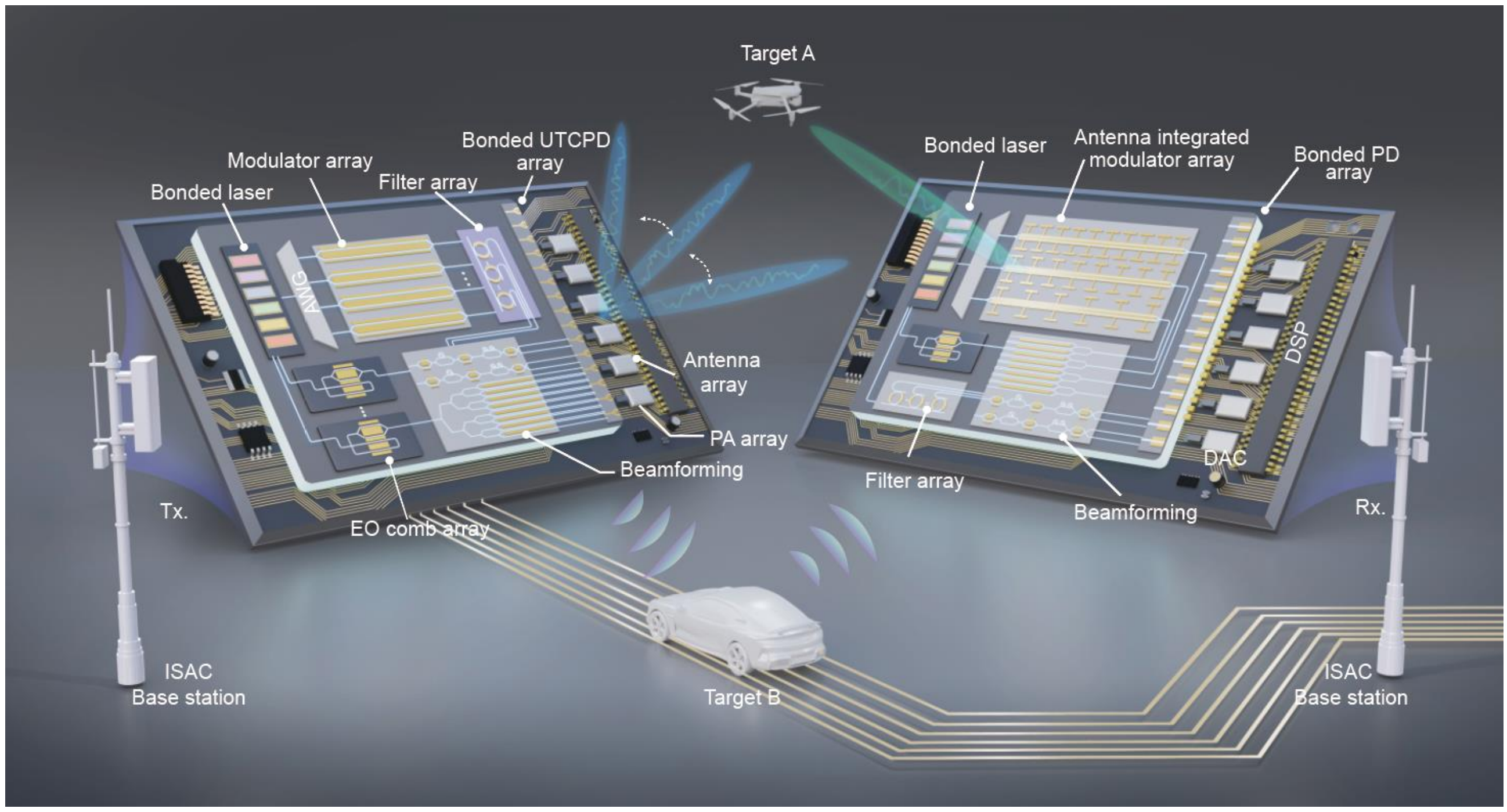

**Fig. 6. Envisioned fully integrated and multi-stream TFLN CPO transceivers for 6G ISAC system.** Tx., transmitter; Rx., receiver.

## Acknowledgment.

This work is supported by the Research Grants Council, University Grants Committee (STG3/E-704/23-N, JRFS2526-1S01), Guangdong and Hong Kong Universities '1+1+1' Joint Research Collaboration Scheme.

## Data availability.

Data underlying the results presented in this paper are not publicly available at this time but may be obtained from the authors upon reasonable request.

## References

1 Dang, S., Amin, O., Shihada, B. & Alouini, M.-S. What should 6G be? Nature Electronics 3, 20-29 (2020).

2 Yang, P., Xiao, Y., Xiao, M. & Li, S. 6G wireless communications: Vision and potential techniques. IEEE network 33, 70-75 (2019).

3 Akyildiz, I. F., Kak, A. & Nie, S. 6G and beyond: The future of wireless communications systems. IEEE access 8, 133995-134030 (2020).

4 Wang, W. et al. On-chip topological beamformer for multi-link terahertz 6G to XG wireless. Nature 632, 522-527 (2024).

5 Govind, B., Anderson, M. G., Wu, F. O., McMahon, P. L. & Apsel, A. An integrated microwave neural network for broadband computation and communication. Nature Electronics, 1-13 (2025).

6 Gao, L. & Rebeiz, G. M. Wideband bandpass filter for 5G millimeter-wave application in 45-nm CMOS silicon-on-insulator. IEEE Electron Device Letters 42, 1244-1247 (2021).

7 Luo, Y.-L. et al. A 16–32 GHz RF Photonic Front-End With 22 nm CMOS Driver and Silicon Travelling-Wave Mach-Zehnder Modulator. Journal of Lightwave Technology 42, 8127-8136 (2024).

8 Zhu, M. et al. Radio-over-fiber access architecture for integrated broadband wireless services. Journal of Lightwave Technology 31, 3614-3620 (2013).

9 Capmany, J. & Novak, D. Microwave photonics combines two worlds. Nature photonics 1, 319 (2007).

10 Yao, J., Capmany, J. & Li, M. Microwave Photonics. (John Wiley & Sons, 2024).

11 Marpaung, D., Yao, J. & Capmany, J. Integrated microwave photonics. Nature photonics 13, 80-90 (2019).

12 Shu, H. et al. Microcomb-driven silicon photonic systems. Nature 605, 457-463 (2022).

13 Pérez, D. et al. Multipurpose silicon photonics signal processor core. Nature communications 8, 636 (2017).

14 Zhuang, L., Roeloffzen, C. G., Hoekman, M., Boller, K.-J. & Lowery, A. J. Programmable photonic signal processor chip for radiofrequency applications. Optica 2, 854-859 (2015).

15 Roeloffzen, C. G. et al. Silicon nitride microwave photonic circuits. Optics express 21, 22937-22961 (2013).

16 Fandiño, J. S., Muñoz, P., Doménech, D. & Capmany, J. A monolithic integrated photonic microwave filter. Nature Photonics 11, 124-129 (2017).

17 Liu, W. et al. A fully reconfigurable photonic integrated signal processor. Nature Photonics 10, 190-195 (2016).

18 Wang, L., Wang, X. & Pan, S. Microwave photonics empowered integrated sensing and communication for 6G. IEEE Transactions on Microwave Theory and Techniques (2025).

19 Rahim, A. et al. Taking silicon photonics modulators to a higher performance level: state-of-the-art and a review of new technologies. Advanced Photonics 3, 024003-024003 (2021).

20 Zhu, D. et al. Integrated photonics on thin-film lithium niobate. Advances in Optics and Photonics 13, 242-352 (2021).

21 Boes, A. et al. Lithium niobate photonics: Unlocking the electromagnetic spectrum. Science 379, eabj4396 (2023).

22 Zhang, M., Wang, C., Kharel, P., Zhu, D. & Lončar, M. Integrated lithium niobate electro-optic modulators: when performance meets scalability. Optica 8, 652-667 (2021).

23 Wang, C. et al. Integrated lithium niobate electro-optic modulators operating at CMOS-compatible voltages. Nature 562, 101-104 (2018).

24 Xu, M. et al. Dual-polarization thin-film lithium niobate in-phase quadrature modulators for terabit-per-second transmission. Optica 9, 61-62 (2022).

25 Kharel, P., Reimer, C., Luke, K., He, L. & Zhang, M. Breaking voltage–bandwidth limits in integrated lithium niobate modulators using micro-structured electrodes. Optica 8, 357-363 (2021).
26 Liu, H. et al. Ultra-High-Efficiency Dual-Band Thin-Film Lithium Niobate Modulator Incorporating Low-k Underfill with 220 GHz Extrapolated Bandwidth for 390 Gbit/s PAM8 Transmission. arXiv preprint arXiv:2411.15037 (2024).
27 Zhu, X. et al. Capacitive-loaded traveling wave electrodes on thin film lithium niobate for sub-terahertz operation. Optical Materials Express 15, 513-521 (2025).
28 Zhang, Y. et al. Monolithic lithium niobate photonic chip for efficient terahertz-optic modulation and terahertz generation. arXiv preprint arXiv:2406.19620 (2024).
29 Zhu, X. et al. Folded Sub-1V V π Thin Film Lithium Niobate Phase Modulator. IEEE Photonics Technology Letters (2025).
30 Zhang, Y. et al. Systematic investigation of millimeter-wave optic modulation performance in thin-film lithium niobate. Photonics Research 10, 2380-2387 (2022).
31 Xue, X. et al. High-performance thin-film lithium niobate Mach-Zehnder modulator on thick silica buffering layer. arXiv preprint arXiv:2412.12556 (2024).
32 Chen, G. et al. Compact 100GBaud driverless thin-film lithium niobate modulator on a silicon substrate. Optics express 30, 25308-25317 (2022).
33 Della Torre, A. et al. Folded electro-optical modulators operating at CMOS voltage level in a thin-film lithium niobate foundry process. Optics Express 33, 6747-6757 (2025).
34 Xu, M., Gao, S., Tan, H. & Cai, X. in Optical Fiber Communication Conference. Th1J. 3 (Optica Publishing Group).
35 Zhang, M., Wang, C., Cheng, R., Shams-Ansari, A. & Lončar, M. Monolithic ultra-high-Q lithium niobate microring resonator. Optica 4, 1536-1537 (2017).
36 Gao, R. et al. Lithium niobate microring with ultra-high Q factor above 10^ 8. Chinese Optics Letters 20, 011902 (2022).
37 Xie, Y. et al. Ultra-Compact and High-Speed Thin-Film Lithium Niobate Tunable Optical Delay Lines. Laser & Photonics Reviews, e01757
38 Feng, H. et al. On-chip optical vector analysis based on thin-film lithium niobate single-sideband modulators. Advanced Photonics 6, 066006-066006 (2024).
39 Zhang, M. et al. Broadband electro-optic frequency comb generation in a lithium niobate microring resonator. Nature 568, 373-377 (2019).
40 Hu, Y. et al. High-efficiency and broadband on-chip electro-optic frequency comb generators. Nature photonics 16, 679-685 (2022).
41 Song, Y., Hu, Y., Lončar, M. & Yang, K. Hybrid Kerr-electro-optic frequency combs on thin-film lithium niobate. Light: Science & Applications 14, 270 (2025).
42 Lv, X. et al. Broadband microwave-rate dark pulse microcombs in dissipation-engineered LiNbO3 microresonators. Nature Communications 16, 2389 (2025).
43 Nie, B. et al. Soliton microcombs in X-cut LiNbO3 microresonators. eLight 5, 15 (2025).
44 Zhang, K. et al. A power-efficient integrated lithium niobate electro-optic comb generator. Communications Physics 6, 17 (2023).
45 Chen, Z. et al. Microwave-resonator-enabled broadband on-chip electro-optic frequency comb generation. Photonics Research 13, 426-432 (2025).
46 Wang, H. et al. in 2023 Optical Fiber Communications Conference and Exhibition (OFC). 1-3 (IEEE).

47 He, L. et al. Low-loss fiber-to-chip interface for lithium niobate photonic integrated circuits. Optics letters 44, 2314-2317 (2019).
48 Ying, P. et al. Low-loss edge-coupling thin-film lithium niobate modulator with an efficient phase shifter. Optics letters 46, 1478-1481 (2021).
49 Rao, A. et al. High-performance and linear thin-film lithium niobate Mach–Zehnder modulators on silicon up to 50 GHz. Optics letters 41, 5700-5703 (2016).
50 He, M. et al. High-performance hybrid silicon and lithium niobate Mach–Zehnder modulators for 100 Gbit s− 1 and beyond. Nature photonics 13, 359-364 (2019).
51 Feng, H. et al. Ultra-high-linearity integrated lithium niobate electro-optic modulators. Photonics Research 10, 2366-2373 (2022).
52 Xu, H. et al. High-frequency and high-linearity lithium niobate electro-optic modulator. ACS Photonics 11, 3232-3238 (2024).
53 Li, D. et al. Compact thin-film lithium niobate dual-stage intensity modulator with ultrahigh linearity. Optics Letters 50, 3329-3332 (2025).
54 Qin, Y.-Y. et al. Broadband, high-linearity TFLN electro-optic modulator for integrated microwave photonics using reconfigurable dual-parallel modulation. Optics Express 33, 13805-13815 (2025).
55 Wei, C. et al. Programmable multifunctional integrated microwave photonic circuit on thin-film lithium niobate. Nature communications 16, 2281 (2025).
56 Xie, X. et al. Broadband millimeter-wave frequency mixer based on thin-film lithium niobate photonics. Electromagnetic Science 3, 0090462-0090461-0090462-0090410 (2025).
57 Maleki, L. The optoelectronic oscillator. Nature Photonics 5, 728-730 (2011).
58 Hao, T. et al. Recent advances in optoelectronic oscillators. Advanced Photonics 2, 044001-044001 (2020).
59 Tao, Z. et al. Ultrabroadband on-chip photonics for full-spectrum wireless communications. Nature, 1-8 (2025).
60 Guo, X. et al. High-performance modified uni-traveling carrier photodiode integrated on a thin-film lithium niobate platform. Photonics Research 10, 1338-1343 (2022).
61 Xie, X. et al. A 3.584 Tbps coherent receiver chip on InP-LiNbO3 wafer-level integration platform. Light: Science & Applications 14, 172 (2025).
62 Wang, L. et al. in 2025 Optical Fiber Communications Conference and Exhibition (OFC). 1-3 (IEEE).
63 Wei, C. et al. Ultra-wideband waveguide-coupled photodiodes heterogeneously integrated on a thin-film lithium niobate platform. arXiv preprint arXiv:2305.07861 (2023).
64 Hao, P. et al. Fourier domain mode-locked optoelectronic oscillator with an electrically tuned thin-film lithium niobate micro-ring filter. Photonics Research 13, 1964-1972 (2025).
65 Han, Z. et al. Integrated Ultra-Wideband Tunable Fourier Domain Mode-Locked Optoelectronic Oscillator. Laser & Photonics Reviews 19, 2402094 (2025).
66 Zeng, Z. et al. Microwave pulse generation via employing an electric signal modulator to achieve time-domain mode locking in an optoelectronic oscillator. Optics letters 46, 2107-2110 (2021).
67 Ma, R. et al. V-band ultra-fast tunable thin-film lithium niobate Fourier-domain mode-locked optoelectronic oscillator. Light: Science & Applications 14, 398 (2025).
68 Lampert, Y. et al. Photonics-integrated terahertz transmission lines. Nature Communications 16, 7004 (2025).
69 Gaier, A. et al. Antenna-coupled integrated millimeterwave modulators and resonant electro-optic frequency combs. arXiv preprint arXiv:2505.04585 (2025).

70 Moller de Freitas, M. et al. Monolithically integrated ultra-wideband photonic receiver on thin film lithium niobate. Communications Engineering 4, 55 (2025).
71 Juneghani, F. A. et al. Wireless microwave-to-optical conversion on thin-film lithium niobate. Journal of Lightwave Technology 42, 5583-5590 (2024).
72 Qian, Y. et al. A D-Band Highly Integrated Low-Noise I/Q Receiver With On-Chip Antenna and Lossy Embedding Network in 40-nm CMOS. IEEE Transactions on Microwave Theory and Techniques (2025).
73 Thyagarajan, S. V., Kang, S. & Niknejad, A. M. A 240 GHz fully integrated wideband QPSK receiver in 65 nm CMOS. IEEE Journal of Solid-State Circuits 50, 2268-2280 (2015).
74 Guo, H. et al. A 460-GHz Receiver Using Second-Order Subharmonic Mixer in 65-nm CMOS. IEEE Transactions on Microwave Theory and Techniques (2024).
75 Feng, H. et al. Integrated lithium niobate microwave photonic processing engine. Nature 627, 80-87 (2024).
76 Xie, Y. et al. Ultra-wideband tunable thin-film lithium-niobate-on-insulator microwave photonic filter. ACS Photonics 12, 1689-1697 (2025).
77 Yan, Y., Yue, H., Xiong, B. & Chu, T. Wideband tunable integrated microwave photonic filter with dual output on thin-film lithium niobate. Optics Letters 50, 5322-5325 (2025).
78 Zhu, S. et al. Integrated lithium niobate photonic millimetre-wave radar. Nature Photonics 19, 204-211 (2025).
79 Ye, K. et al. Integrated Brillouin photonics in thin-film lithium niobate. Science Advances 11, eadv4022 (2025).
80 Rodrigues, C. C. et al. Cross-polarized stimulated Brillouin scattering in lithium niobate waveguides. Physical review letters 134, 113601 (2025).
81 Tao, Y. et al. Integrated photonic ultrawideband real-time spectrum sensing for 6G wireless networks. arXiv preprint arXiv:2509.03874 (2025).
82 Yan, H. et al. Thin-Film-Lithium-Niobate Photonic Chip for Ultra-Wideband and High-Precision Microwave Frequency Measurement. Laser & Photonics Reviews 19, 2401273 (2025).
83 Wang, L. et al. Integrated Ultra-Wideband Dynamic Microwave Frequency Identification System in Lithium Niobate on Insulator. Laser & Photonics Reviews 18, 2400332 (2024).
84 Feng, H. et al. in CLEO: Applications and Technology. AA119_117 (Optica Publishing Group).
85 Xie, Y. et al. Ultra-Compact and High-Speed Thin-Film Lithium Niobate Tunable Optical Delay Lines. Laser & Photonics Reviews 19, e01757 (2025).
86 Ke, W. et al. Digitally tunable optical delay line based on thin-film lithium niobate featuring high switching speed and low optical loss. Photonics Research 10, 2575-2583 (2022).
87 Qin, Y. Y. et al. Broadband Lithium Niobate Integrated Microwave Photonic Beamforming Chip. Laser & Photonics Reviews, e00219 (2025).